\documentclass[11pt,twoside]{article}
\usepackage{subfigure}
\usepackage{epsfig}
\def\Journal#1#2#3#4{{#1} {\bf #2}, #3 (#4)}

\def\NPB{{\em Nucl. Phys.} B}
\def\PLB{{\em Phys. Lett.}  B}
\def\PRL{\em Phys. Rev. Lett.}
\def\PRD{{\em Phys. Rev.} D}

\newcommand{\MET}{\hbox{$\rlap{\,/}E_T $}}

\begin{document}
\begin{flushright}
\large
FERMILAB-Conf-97/206-E
\end{flushright}
\vskip 0.3in
\begin{center}
\Large
{\bf Tevatron Results on Gauge Boson Couplings}
\end{center}
\begin{center}
\vskip 0.3in
T.~Yasuda\\
{\em Northeastern University, Boston, MA 02115\\}
\vskip 0.15in
for the CDF and D{\O} Collaborations\\
\vskip 0.3in
\end{center}

\begin{abstract}

Direct measurements of the trilinear gauge boson
couplings by the CDF and D{\O} collaborations at Fermilab are reviewed.
Limits on the anomalous couplings were obtained at a $95\%$ CL
from four diboson production processes:
$W\gamma$ production with the subsequent $W$ boson decay to $e\nu$ or $\mu\nu$,
$WW$ production with both the $W$ boson decaying to $e\nu$ or $\mu\nu$,
$WW/WZ$ production with one of the gauge bosons decaying leptonically and
the other gauge boson decaying to two jets, 
and $Z\gamma$ production
with the subsequent $Z$ boson decay to $ee$, $\mu\mu$, or $\nu\nu$.
Limits were also obtained by a combined fit to $W\gamma$, $WW\rightarrow$
dileptons and $WW/WZ$ semileptonic data samples.

\end{abstract}

\section*{Introduction}

The gauge boson self-interactions are a direct consequence of the
non-Abelian $SU(2)\times U(1)$ gauge symmetry of the Standard Model (SM).
The trilinear gauge boson couplings can be
measured by studying the gauge boson pair production processes.
The determination of the couplings is one of a few remaining crucial
tests of the SM.
Deviations of the couplings from the SM values signal new physics.
Measurements of the anomalous couplings have been previously reported by the
UA2\cite{UA2}, CDF\cite{Wg1a_cdf},\cite{WWWZ1a_cdf},\cite{Zg1a_cdf}
and D{\O}\cite{Wg1a_d0},\cite{WW1a_d0},\cite{WWWZ1a_d0},\cite{Zg1a_d0}
collaborations.

The $WWV (V = \gamma ~{\rm or}~ Z)$ vertices are described by a
general effective Lagrangian\cite{Wcoupling}
with two overall couplings ($g_{WW\gamma} = -e$ and
$g_{WWZ} = -e \cdot \cot \theta_{W}$) and
six dimensionless couplings
$g_{1}^{V}$, $\kappa_V$ and $\lambda_V$, where $V = \gamma$ or $Z$,
after imposing {\it C}, {\it P} and {\it CP} invariance.
$g_{1}^{\gamma}$ is
restricted to unity by electromagnetic gauge invariance.
The SM Lagrangian is obtained by setting $g_1^{\gamma} = g_1^Z = 1$,
 $\kappa_{V} = 1 (\Delta\kappa_{V} \equiv \kappa_V - 1 = 0)$ and
 $\lambda_V = 0$.
The cross section with the non-SM couplings grows with
${\hat s}$.
In order to avoid unitarity violation,
the anomalous couplings are modified as form factors with a scale $\Lambda$;
 $\lambda_{V}(\hat{s}) = {\lambda_{V} \over ( 1 + \hat{s}/\Lambda^{2})^{2}}$ and
 $\Delta\kappa_{V}(\hat{s}) = 
{\Delta\kappa_{V} \over ( 1 + \hat{s}/\Lambda^{2})^{2}}$.

The $Z\gamma V(V = \gamma ~{\rm or}~ Z)$ vertices are described by
a general vertex function\cite{Zcoupling}
with eight dimensionless couplings
$h_{i}^{V} (i=1,4 ~; V=\gamma ~{\rm or}~ Z)$.
In the SM, all of $h_{i}^{V}$'s are zero.
The form factors for these couplings,
similar to the $WWV$ couplings, are
 $h_{i}^{V}(\hat{s}) = {h_{i0}^{V} \over ( 1 + \hat{s}/\Lambda^{2})^{n}}$,
where $n = 3$ for $i = 1,3$ and $n = 4$ for $i = 2,4$.

The characteristic that the production cross section of a gauge boson pair with
anomalous couplings grows with ${\hat s}$ is an advantage for the Tevatron
experiments over LEP II.
The increase of the cross section is greater at higher gauge boson $p_T$.
This is exploited to set limits on the anomalous couplings in all of the
analyses presented here.

In this report, the measurements of trilinear gauge boson couplings by the
CDF and D{\O} experiments at Fermilab are reviewed. 
Limits on the anomolous couplings
were obtained at a $95\%$ CL
from four processes:
$W\gamma$ production with the subsequent $W$ boson decay to $e\nu$ or $\mu\nu$,
$WW$ production with both the $W$ boson decaying to $e\nu$ or $\mu\nu$,
$WW/WZ$ production with one of the gauge bosons decaying leptonically and
the other gauge boson decaying to two jets, 
and $Z\gamma$ production
with the subsequent $Z$ boson decay to $ee$, $\mu\mu$, or $\nu\nu$.
Limits were also obtained by a combined fit to $W\gamma$, $WW\rightarrow$
dileptons and $WW/WZ$ semileptonic data samples.

\section*{$W\gamma$ production}

The $W(\ell\nu)\gamma$ candidates were selected by
searching for events containing an isolated electron or muon with high 
transverse energy, $E_T$, large
missing transverse energy, \MET, and an isolated photon.
The major sources of background for this process are $W +$ jets production with a jet
misidentified as a photon and $Z\gamma$ production with an electron
or a muon from $Z$ decay undetected.
The signal to background ratio for this process is 1 to 0.2 -- 0.3 for both
CDF and D{\O} experiments.
CDF and D{\O} previously reported the results with the 1992 -- 1993
data \cite{Wg1a_cdf}, \cite{Wg1a_d0}.

CDF has reported the preliminary results based on a partial data set of
1993 -- 1995 Tevatron collider run ($\sim 67 ~{\rm pb}^{-1}$)
\cite{CDF_WgZg_1b}.
The candidate events were required to have an electron or a muon with $E_T > 20
~{\rm GeV}$, a photon with $E_T > 7$ GeV, and $\MET > 20$ GeV.
A requirement on the transverse mass $M_T > 40 ~{\rm GeV/c^{2}}$ was applied
to insure the detection of a W boson.
The electrons and photons had to be in the fiducial region of $|\eta| < 1.1$
and the muons in $|\eta| < 0.6$.
In addition, the separation in $\eta-\phi$ space between a photon and a lepton
(${\cal R}_{\ell\gamma}$) had to be greater than 0.7.
This requirement suppressed the contribution of the radiative $W$ decay
process, and minimized the probability for a photon cluster to merge with a
nearby calorimeter cluster associated with an electron or a muon.
A total of 75 $e\nu\gamma$ and 34 $\mu\nu\gamma$ candidate events was observed.

D{\O} completed the analysis of the data set of the 1993 -- 1995 run,
and the results from the two Tevatron runs were combined.
The total data sample corresponds to
an integrated luminosity of $92.8~{\rm pb}^{-1}$
\cite{Wg1b_d0}.
For the electron channel, the candidate events were required to have an
electron with
$E_{T} > 25 ~{\rm GeV}$ in the fiducial region of $|\eta| < 1.1$ or
$1.5 < |\eta| < 2.5$ and to have $\MET > 25 ~{\rm GeV}$.
A requirement on the transverse mass $M_T > 40 ~{\rm GeV/c^{2}}$ was also
applied.
For the muon channel, the events were required to have a muon
with $p_T > 15 ~{\rm GeV/c}$ in the fiducial region of $|\eta| < 1.0$ and
to have $\MET > 15 ~{\rm GeV}$.
The requirement for the photon was common to both the channels.
The candidate events were required to have a photon with 
$E_{T} > 10 ~{\rm GeV}$ in the fiducial region
of $|\eta| < 1.1$ or $1.5 < |\eta| < 2.5$.
The same cut as CDF on the separation between the lepton and the photon was
applied.
The above selection criteria yielded 57 $W(e\nu)\gamma$ and 70 $W(\mu\nu)\gamma$
candidates.

The backgrounds were estimated from Monte Carlo simulation and data.
The estimated total backgrounds are listed in Table~\ref{table:Wg}.
The detection efficiency was estimated
as a function of anomalous couplings
using the Monte Carlo program of Baur and Zeppenfeld\cite{Zeppnfld} and
a fast detector simulation program.
The $W\gamma$ cross section times the leptonic branching ratio
$Br(W\rightarrow\ell\nu)$
(for photons with $E_{T}^{\gamma}>7 ~{\rm GeV}$ (CDF) or $> 10 ~{\rm GeV}$
({D{\O}) and ${\cal R}_{\ell\gamma}>0.7$) was obtained from the number of
candidate events and the estimated numbers of background events, as listed in
Table~\ref{table:Wg}.
\begin{table}[htb]
\begin{center}
\begin{tabular}{ccccc}
&
\multicolumn{2}{c}{D{\O} ($\int {\cal L}dt = 92.8~{\rm pb}^{-1}$)}&
\multicolumn{2}{c}{CDF ($\int {\cal L}dt \sim 67~{\rm pb}^{-1}$)} \\ \hline
&
$e\nu\gamma$& $\mu\nu\gamma$& $e\nu\gamma$& $\mu\nu\gamma$\\ \hline
${\rm N}_{\rm data}$& 57& 70& 75& 34\\
${\rm N}_{\rm BG}$& $15.2 \pm 2.5$& $27.7 \pm 4.7$&
$16.1 \pm 2.4$& $10.3 \pm 1.2$\\
${\rm N}_{\rm Signal}$& $41.8 \pm 8.9$& $42.3 \pm 9.7$&
$58.9 \pm 9.4$& $23.7 \pm 6.0$\\
$\sigma\cdot {\rm BR}$&
\multicolumn{2}{c}{$11.3_{-1.5}^{+1.7} \pm 1.4 \pm 0.6 ~{\rm pb}$}&
\multicolumn{2}{c}{$20.7 \pm 2.9 \pm 0.7 ~{\rm pb}$}\\
$\sigma\cdot {\rm BR} ({\rm SM})$&
\multicolumn{2}{c}{$12.5 \pm 1.0 ~{\rm pb}$}&
\multicolumn{2}{c}{$18.6 \pm 2.9 ~{\rm pb}$}\\
\end{tabular}
\caption{Summary of $W\gamma$ analyses}
\label{table:Wg}
\end{center}
\end{table}
The results from the two experiments agree with the SM prediction within errors.

To set limits on the anomalous couplings, a binned maximum
likelihood fit was performed on the $E_T$ spectrum of the photon.
Form factors with a scale $\Lambda=1.5 ~{\rm TeV}$ were used in the
Monte Carlo event generation.
The $95\%$ CL limit contour for the CP-conserving anomalous couplings
$\Delta\kappa_{\gamma}$ and $\lambda_{\gamma}$ by D{\O} are shown in
Fig.~\ref{fig:Wg_contours}.
The SM point and the point for the models with $U(1)$ couplings only are
also indicated in Fig.~\ref{fig:Wg_contours}.
The $95\%$ CL limits on the anomalous couplings are listed in
Table~\ref{table:Wg_limits}.
\begin{table}[htb]
\begin{center}
\begin{tabular}{ccc}
& $\lambda_{\gamma} = 0$& $\Delta\kappa_{\gamma} = 0$\\ \hline 
D{\O}& $-0.93 < \Delta\kappa_{\gamma} < 0.94$&
$-0.31 < \lambda_{\gamma} < 0.29$ \\
CDF (preliminary)& $-1.8 < \Delta\kappa_{\gamma} < 2.0$&
$-0.7 < \lambda_{\gamma} < 0.6$ \\
\end{tabular}
\caption{$95~\%$ CL limits on the anomalous $WW\gamma$ couplings}
\label{table:Wg_limits}
\end{center}
\end{table}
The $U(1)$ only couplings of the $W$ boson to a photon, which correspond to
$\kappa_{\gamma}=0 ~(\Delta\kappa_{\gamma}=-1)$ and $\lambda_{\gamma}=0$
are excluded at a $96\%$ CL by the D{\O} results.
The D{\O} limit on $\lambda_{\gamma}$ is the tightest to date among
the individual analyses of gauge boson pair final states.

\begin{figure}[htb]
 \centerline{\includegraphics[clip,width=12cm,bb = 0 400 567 712]
 {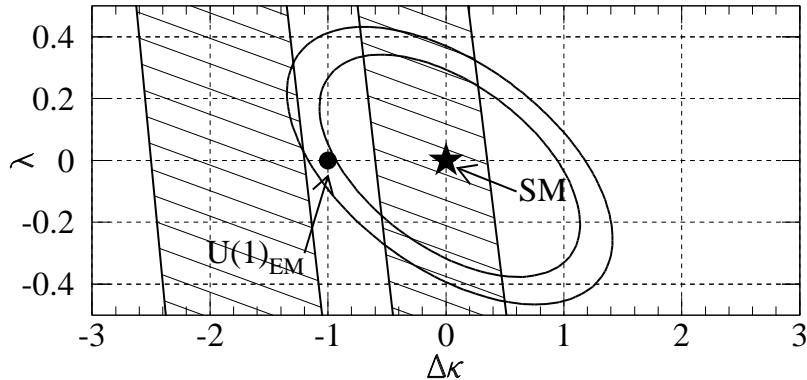}}
 \caption{$95~\%$ CL limits on the anomalous $WW\gamma$ couplings from the D{\O}
 $W\gamma$ analysis. The shaded bands are the constraints from CLEO.}
 \label{fig:Wg_contours}
\end{figure}

\section*{$WW\rightarrow$ dileptons}

The $W$ boson pair production candidates were obtained by searching for
events containing two isolated leptons ($e\mu, ~ee, ~{\rm or}~ \mu\mu$)
with high $E_{T}$ and large $\MET$.
The major sources of background for this process are Drell-Yan production of
a $Z$ boson or
a virtual photon, $t{\bar t}$ production, $W\gamma$ production with a $\gamma$
misidentified as an electron, $Z\rightarrow\tau\tau$ with the subsequent
$\tau$ decays to $e\nu\nu$ or $\mu\nu\nu$, and $W +$ jets production with
a jet misidentified as an electron or a muon.
The signal to background ratio for this process is 1 to 0.5(1) for CDF(D{\O}).
D{\O} previously reported the results with the 1992 -- 1993 data
\cite{WW1a_d0}.

CDF analysed the full data set of 1992 -- 1993 and 1993 -- 1995 Tevatron
collider runs, corresponding to an integrated luminosity of
$108~{\rm pb}^{-1}$ \cite{WW1b_cdf}.
The candidate events were required to have two electrons, two muons, or one
electron and one muon with each having $E_T > 20$ GeV and $\MET > 20$ GeV.
An invariant mass cut, $75 < m < 105~{\rm GeV/c^2}$, was applied to $ee$ and $\mu\mu$
candidate events, in order to reduce the background from the $Z$ boson
production.
The background contribution from the $t{\bar t}$ produciton was suppressed by
rejecting the events with one or more jets with $E_T > 10$ GeV.
CDF observed five candidate events with an estimated background of
$1.2 \pm 0.3$ events, as listed in Table~\ref{table:WW_dilep}.
From this, the $WW$ production cross section was calculated to be
\begin{eqnarray*}
\sigma({\bar p}p\rightarrow W^+W^-) &=& 10.2_{-5.1}^{+6.3}~({\rm stat}) \pm
1.6~({\rm syst})~{\rm pb}.
\end{eqnarray*}
The probability that the observed events corresponds to a fluctuation of
the background is $1.1 \%$.
The measured cross section is consistent with the SM prediction of
$9.5~{\rm pb}$. 
\begin{table}[htb]
\begin{center}
\begin{tabular}{ccc}
& D{\O} ($\int {\cal L}dt = 96.6~{\rm pb}^{-1}$)&
CDF ($\int {\cal L}dt = 108~{\rm pb}^{-1}$)\\ \hline
${\rm N}_{\rm data}$& 5& 5\\
${\rm N}_{\rm BG}$& $3.3 \pm 0.4$& $1.2 \pm 0.3$\\ 
${\rm N}_{\rm SM}$& $2.10 \pm 0.15$& $3.5 \pm 1.2$\\
\end{tabular}
\caption{Summary of $WW\rightarrow$ dileptons analyses}
\label{table:WW_dilep}
\end{center}
\end{table}

D{\O} also completed the analysis of the data set of the
1993 -- 1995 Tevatron collider run,
and the results from the two Tevatron runs were combined.
The total data sample corresponds to an integrated
luminosity of $96.6~{\rm pb}^{-1}$.
For the $ee$ channel, the candidate events were required to have two electrons,
one with
$E_{T}\geq25 ~{\rm GeV}$ and another with $E_{T}\geq20 ~{\rm GeV}$.
The $\MET$ was required to be $\geq20 ~{\rm GeV}$.
The $Z$ boson background was reduced by removing events with the dielectron
invariant mass between ${\rm 76 ~and~ 106 ~GeV/c^2}$.
For the $e\mu$ channel, an electron with $E_{T}\geq25 ~{\rm GeV}$ and
a muon with $p_{T}\geq15 ~{\rm GeV/c}$ were required.
$\MET$ was required to be $\geq20 ~{\rm GeV}$.
For the $\mu\mu$ channel, two muons were required, one with
$p_{T}\geq25 ~{\rm GeV/c}$ and another with $p_{T}\geq20 ~{\rm GeV/c}$.
In order to reduce the background from the $Z$ boson events, it was required
that the $\MET$ projected on the dimuon bisector in the transverse plane be
greater than 30 GeV.
The $t{\bar t}$ background was suppressed by applying a cut on the hadronic
energy in the event.
It was required that the vector sum of hadronic energy in the event be
$\leq40 ~{\rm GeV}$ in magnitude in all three channels.
For the three channels combined, the expected number of events for
the SM $W$ boson pair production, based on a cross section of 9.5 pb,
is $2.10\pm0.15$.
D{\O} observed five candidate events.
A maximum likelihood fit to the electron $E_T$ and the muon $p_T$ of 
the five candidate events was performed and limits on the anomalous couplings
were obtained.
The preliminary $95 \%$ CL limits from the fit are:
\begin{eqnarray*}
-0.62 < \Delta\kappa < 0.75~(\lambda = 0)\\
-0.50 < \lambda < 0.56~(\Delta\kappa = 0)
\end{eqnarray*}
where the $WW\gamma$ couplings are assumed to be equal to the $WWZ$ couplings
and $\Lambda = 1.5$ TeV is used.  
These limits are comparable to the limits obtained from
the $WW/WZ\rightarrow$ semileptonic mode analyses.

\section*{$WW/WZ\rightarrow\ell\nu jj, \ell\ell jj$}

The WW/WZ candidates were obtained by searching for events containing
an isolated electron or muon with high $E_T$ and large
$\MET$, indicating a $W$ boson decay, or two high $E_T$
electrons or muons, indicating a $Z$ boson decay, and two high $E_T$ jets.
The transverse mass of the electron or muon and neutrino system was required
to be $M_T >40~{\rm GeV/c^2}$ for $e\nu jj$ and $\mu\nu jj$ candidates.
The invariant mass of $ee$ or $\mu\mu$ system was required to be
$70 < m_{\ell\ell} < 110~{\rm GeV/c}^{2}$ for $eejj$ and $\mu\mu jj$ candidates.
There were two major sources of background for this process: QCD multijet
events with a jet misidentified as an electron or a muon and $W$ or $Z$
boson production associated with two jets.
The signal to background ratio for this process is 1 to 20 -- 30
with a low $p_T$ cut on the $W$ or $Z$ boson.
CDF and D{\O} both previously reported the results from the 1992 -- 1993 data
\cite{WWWZ1a_cdf}, \cite{WWWZ1a_d0}.

CDF completed the analysis of the data set of the
1993 -- 1995 Tevatron collider run, and the results from the two
Tevatron collider runs were combined.
The total data sample corresponds to an integrated
luminosity of $110~{\rm pb}^{-1}$ \cite{ichep96}.
The background was eliminated by imposing a high $p_T$ cut
($p_T^{jj} > 200$ GeV) on
the hadronically decaying $W$ or $Z$ boson.
This cut also significantly reduced the signals from the SM $WW/WZ$ production,
but it enabled one to set limits on the anomalous couplings that enhance
the cross section for high $p_T$ gauge bosons.
The invariant mass of the two jet system was required to be
$60 < m_{jj} < 110~{\rm GeV/c}^{2}$, as expected for
a $W$ or $Z$ boson decay.
Before the invariant mass cut on the two jet system, 694 $\ell\nu jj$
and 47 $\ell\ell jj$ candidates were observed.
After the invariant mass cut, no events remained.
The limits on the anomalous couplings from this analysis are listed in
Table~\ref{table:WWWZ_limits}.

D{\O} completed the analysis of $e\nu jj$ channel using
the data set of the 1993 -- 1995 Tevatron collider run,
and the results from the two Tevatron runs were combined.
The total data sample corresponds to an integrated
luminosity of $96~{\rm pb}^{-1}$ \cite{WWWZ1b_d0}.
The candidate events were required to have an electron with $E_T > 25$ GeV,
two or more jets each with $E_T > 20$ GeV and $\MET > 25$ GeV.
The transverse mass of the electron and $\MET$ system was required
to be $M_T >40~{\rm GeV/c^2}$.
The invariant mass of the two jet system was required to be
$50 < m_{jj} < 110~{\rm GeV/c}^{2}$.
The estimated numbers of background events are listed in
Table~\ref{table:WWWZ}.
The SM predicted $20.7 \pm 3.1$ events for the above requirements.
No significant deviation from the SM prediction was seen.
The $p_T$ spectrum of $W$ boson calculated from the $E_T$ of electron and
$\MET$, $p_T^{e\nu}$, which is more precise than the value from
$E_T$ of two jets, is shown in Fig~\ref{fig:WWWZ1b_pt}.
The solid circles and the solid histogram indicate the data and the
background estimate plus SM prediction, respectively.
A maximum likelihood fit to the $p_T^{e\nu}$ spectrum was performed to set
limits on the anomalous couplings.
The limits on the anomalous couplings are listed in
Table~\ref{table:WWWZ_limits}.

\begin{figure}[htb]
 \centerline{\includegraphics[width=12cm]{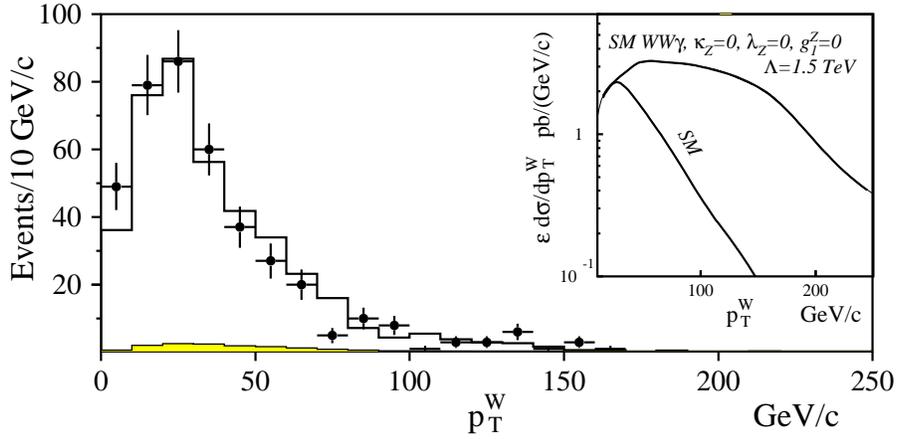}}
 \caption{$p_T$ distribution of the $e\nu$ system from the D{\O} 1993 -- 1995
 data set.}
 \label{fig:WWWZ1b_pt}
\end{figure}

\begin{table}[htb]
\begin{center}
\begin{tabular}{ccc}
D{\O}& $1992-1993$& $1993-1995$\\ \hline
Luminosity& $13.7~{\rm pb}^{-1}$& $82.3~{\rm pb}^{-1}$\\
Backgrounds& &\\
~$W+\ge 2$ jets& $62.2 \pm 13.0$& $279.5 \pm 36.0$\\
~Multijet& $12.2 \pm 2.6$& $104.3 \pm 12.3$\\
~$t{\bar t}\rightarrow e\nu jjX$& $0.9 \pm 0.1$& $3.7 \pm 1.3$\\
~Total& $75.3 \pm 13.3$& $387.5 \pm 38.1$\\
Data& 84& 399\\
SM $WW+WZ$ prediction& $3.2 \pm 0.6$& $17.5 \pm 3.0$\\
\end{tabular}
\caption{Summary of D{\O} $WW/WZ\rightarrow e\nu jj$ analysis}
\label{table:WWWZ}
\end{center}
\end{table}

Different assumptions for the relationship between the $WWZ$ couplings and
the $WW\gamma$ couplings were also examined.
The limit contour in Fig.~\ref{fig:WWWZ1b_contours}b
was obtained using the HISZ\cite{HISZ} relations.
In Figs.~\ref{fig:WWWZ1b_contours}c and ~\ref{fig:WWWZ1b_contours}d
limit contours on the $WWZ$ couplings are shown under the
assumption that the $WW\gamma$ couplings take the SM values.
These plots indicate that this analysis is more sensitive to $WWZ$
couplings as expected from the larger overall couplings for $WWZ$
than $WW\gamma$ and that it is complementary to the $W\gamma$ analysis
which is sensitive to the $WW\gamma$ couplings only.
The $U(1)$ point in the anomalous $WWZ$ couplings plane,
$\kappa_Z = 0 (\Delta\kappa_Z = -1)$, $\lambda_Z = 0$ and
$g_1^Z = 0 (\Delta g_1^Z = -1)$, is excluded at a $99 \%$ CL.
The CDF and D{\O} limits on $\Delta\kappa$ are the tightest limits to date
among the individual analyses of gauge boson pair final states.

\begin{figure}[htb]
 \begin{center}
 \begin{tabular}{cc}
 \subfigure{
 \epsfig{file=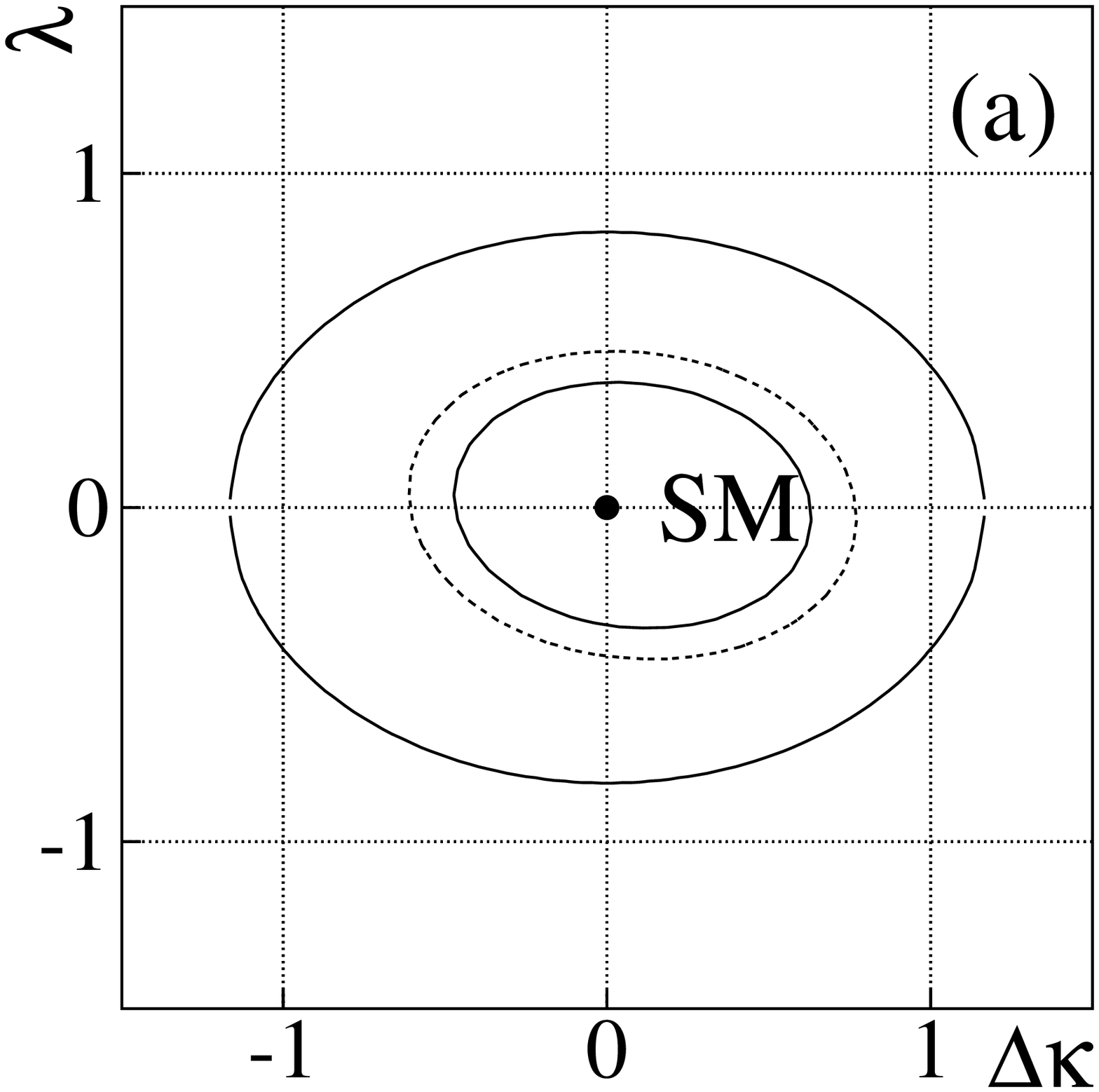,width=2.in}}
 \subfigure{
 \epsfig{file=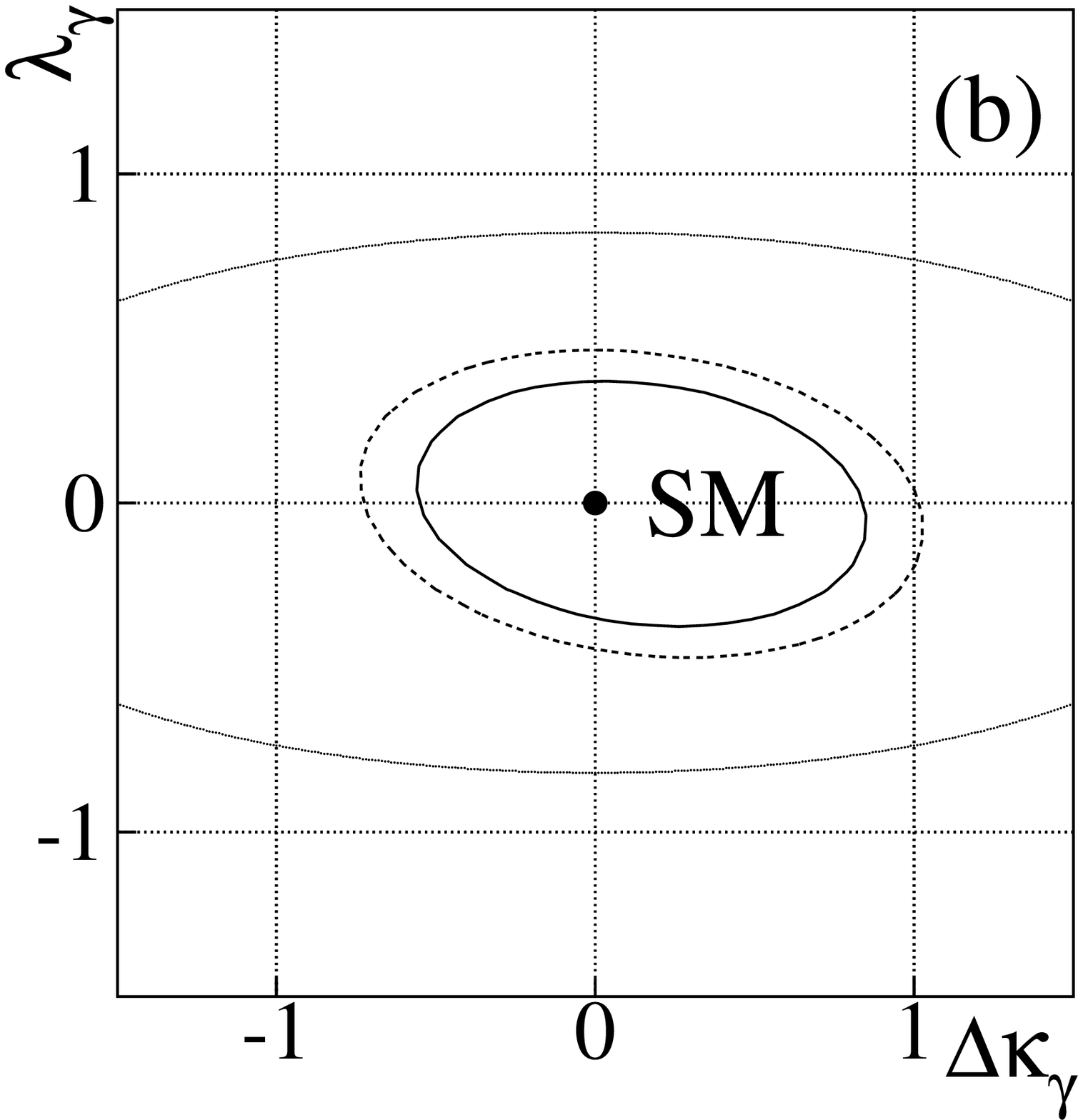,width=2.in}}\\
 \subfigure{
 \epsfig{file=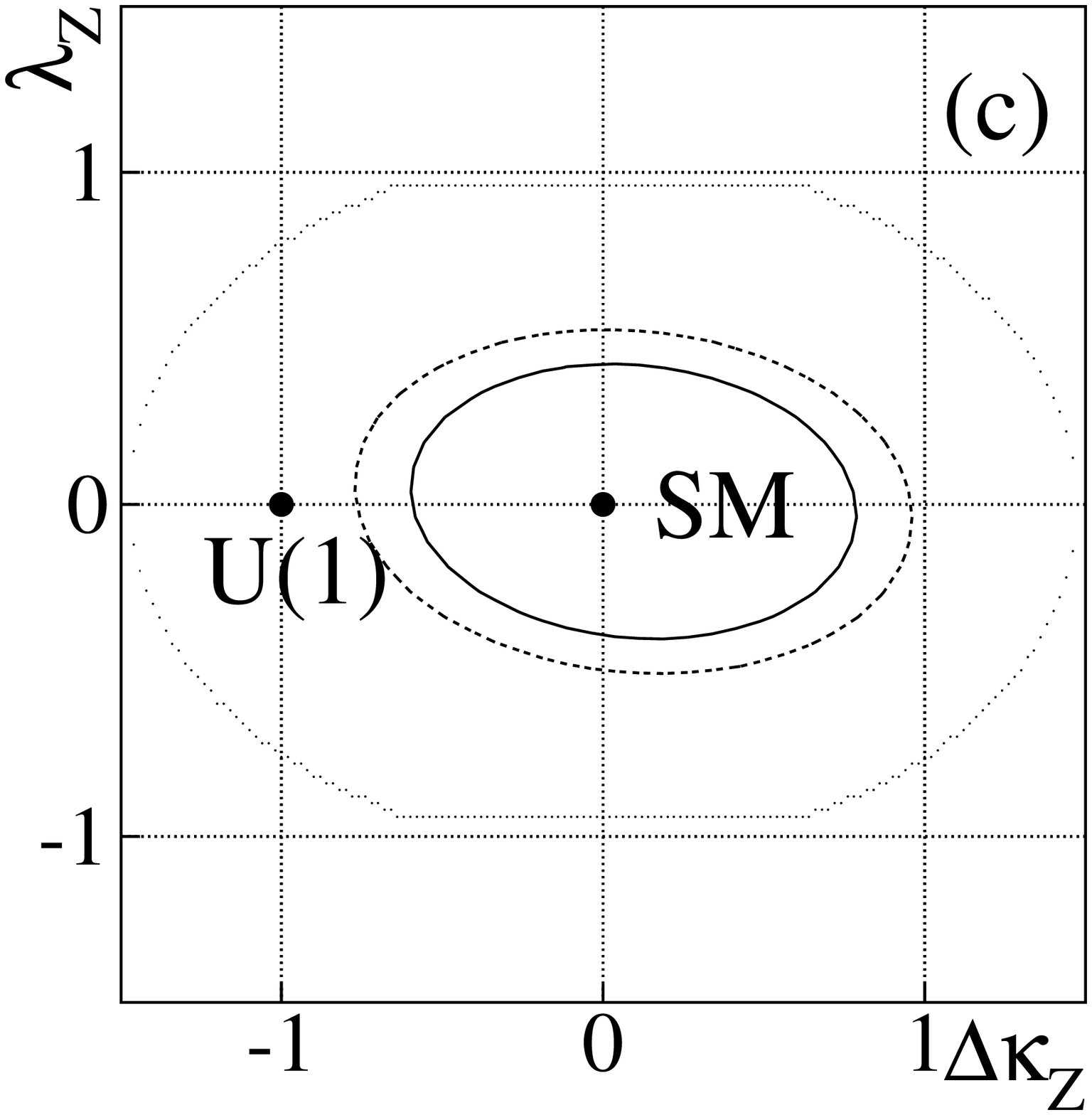,width=2.in}}
 \subfigure{
 \epsfig{file=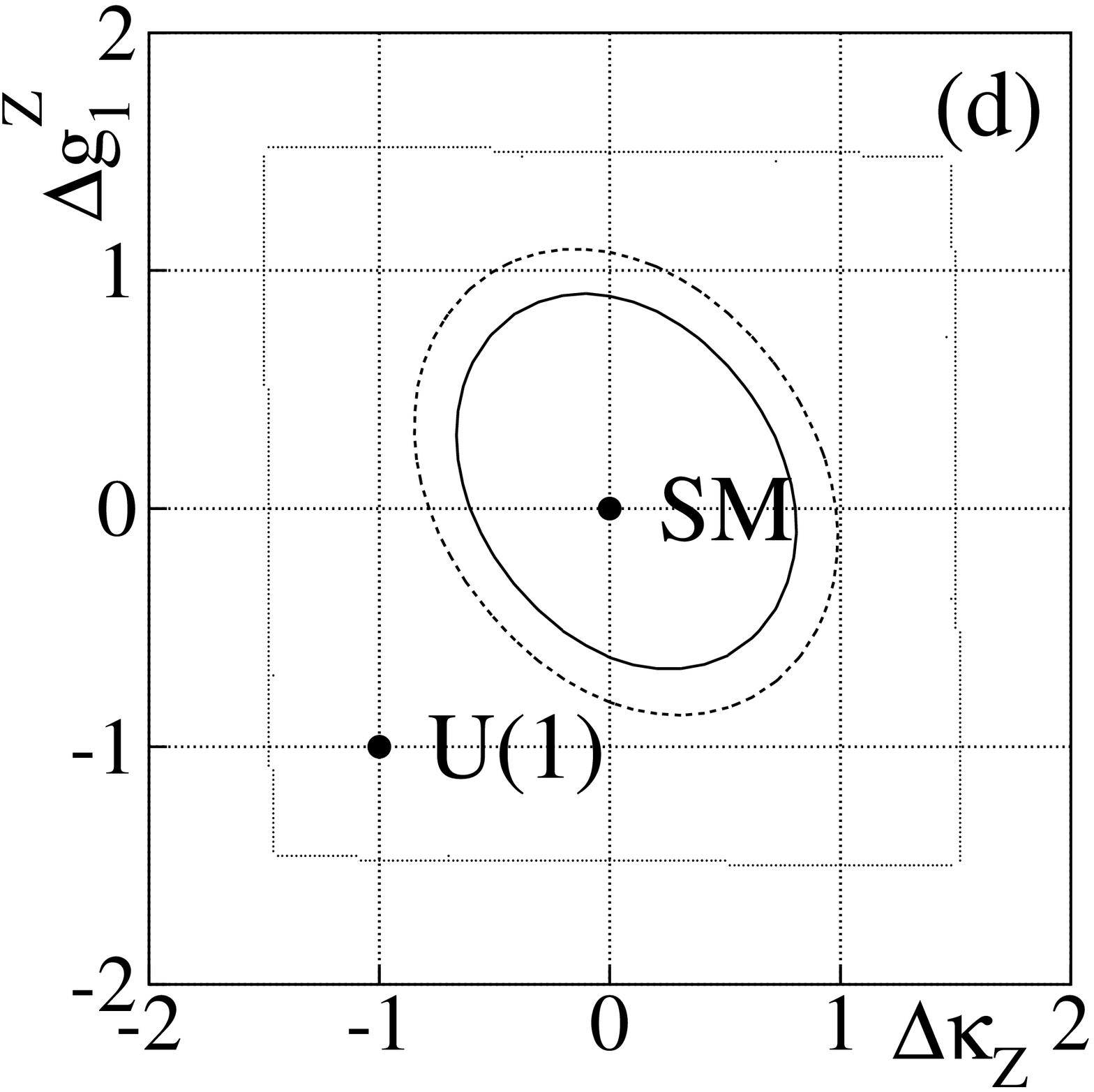,width=2.in}}
 \end{tabular}
 \end{center}
 \caption{D{\O} $95~\%$ CL limits on $CP-$conserving anomalous couplings:
 (a) $\Delta\kappa\equiv\Delta\kappa_{\gamma}=\Delta\kappa_Z,
 ~\lambda\equiv\lambda_{\gamma}=\lambda_Z$;
 (b) HISZ relations;
 (c) and (d) SM $WW\gamma$ couplings.}
 \label{fig:WWWZ1b_contours}
\end{figure}

\begin{table}[htb]
\begin{center}
\begin{tabular}{ccccc}
&
\multicolumn{2}{c}{D{\O}}&
\multicolumn{2}{c}{CDF}\\ \hline
Couplings $\backslash \Lambda$(TeV)& 1.5& 2.0& 1.0& 2.0\\ \hline
$\Delta\kappa_{\gamma} = \Delta\kappa_Z$& -0.47, 0.63&
-0.43, 0.59& -0.67, 0.85& -0.49, 0.54\\
$\lambda_{\gamma} = \lambda_Z$&-0.36, 0.39& -0.33, 0.36&
-0.51, 0.51& -0.35, 0.32\\
$\Delta g_1^Z ({\rm SM}~WW\gamma)$&-0.64, 0.89& -0.60, 0.81&
-0.91, 1.05& -0.61, 0.68\\
$\Delta\kappa_Z ({\rm SM}~WW\gamma)$& -0.60, 0.79& -0.54, 0.72&
-0.95, 1.01& 0.58, 0.68\\
$\lambda_Z ({\rm SM}~WW\gamma)$& -0.40, 0.43& -0.37, 0.40&
-0.60, 0.58& -0.37, 0.40\\
$\Delta\kappa_{\gamma}$ HISZ& -0.56, 0.85& -0.53, 0.78&
-0.83, 1.02& -0.61, 0.67\\
$\lambda_{\gamma}$ HISZ& -0.36, 0.38& -0.34, 0.36&
-0.51, 0.52& -0.34, 0.33\\
\end{tabular}
\caption{Summary of preliminary $95~\%$ CL limits on $WW\gamma$ and $WWZ$
couplings from $WW/WZ$ semileptonic mode analyses}
\label{table:WWWZ_limits}
\end{center}
\end{table}

\section*{Limits on $WW\gamma/WWZ$ couplings from combined fit}

The limits on $WW\gamma$ couplings were obtained from a
fit to the photon $E_T$ spectrum of the $W\gamma$ candidate events.
The limits on $WW\gamma$ and $WWZ$ couplings were obtained
from a fit to the $E_T$ of two leptons of the $WW\rightarrow$ dileptons
candidate events (D{\O}) and a fit to the $p_T$ of electron -- neutrino system
of the $WW/WZ\rightarrow e\nu jj$ candidate events (D{\O}) or
the limit on the cross section with a high $p_T$ cut on the hadronically
decaying gauge bosons (CDF).
Since these analyses measure the
same couplings, D{\O} performed a combined fit to all three data sets
\cite{1a_PRD_d0} from the
1992 -- 1993 and 1993 -- 1995 Tevatron collider runs,
yielding a significantly improved limits from the individual analyses.
The preliminary limits are:
\begin{eqnarray*}
-0.33 < \Delta\kappa < -0.45 ~(\lambda = 0)~;~
-0.20 < \lambda < 0.20 ~(\Delta\kappa = 0),
\end{eqnarray*}
where the $WWZ$ couplings and the $WW\gamma$
couplings are assumed to be equal.
In the fit, correlated uncertainties such as the uncertainties on the
integrated luminosities and theoretical prediction of the cross section were
properly taken into account.
The $95 \%$ CL contour limit is shown in Fig.~\ref{fig:combined_limits}.
These limits represent a significant progress in constraining the
$WW\gamma/WWZ$
couplings in the past several years and are competitive limits to those
expected from the LEP II experiments.

\begin{figure}[htb]
 \centerline{\includegraphics[width=9cm]{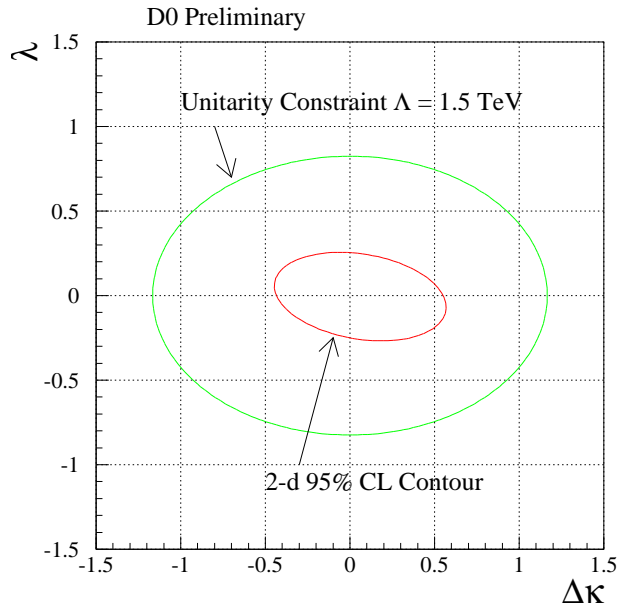}}
 \caption{$95~\%$ CL limits on the anomalous couplings from a combined fit
 to $W\gamma$,
 $WW\rightarrow$ dileptons, and $WW/WZ\rightarrow e\nu jj$ data samples.}
 \label{fig:combined_limits}
\end{figure}

\section*{$Z\gamma$ production}
\subsection*{$Z\gamma\rightarrow ee\gamma, \mu\mu\gamma$}

The $Z\gamma$ candidates were selected by searching for events containing
two isolated electrons or two isolated muons with high $E_T$
and an isolated photon.
The major sources of background for this process are $Z +$ jets production with a jet
misidentified as a photon and multijet and direct photon production events
with two jets misidentified as electrons or muons and a jet misidentified as
a photon.
The signal to background ratio for this process is 1 to 0.1 for both
CDF and D{\O} experiments.
CDF and D{\O} both previously reported the results from the 1992 -- 1993 data
\cite{Zg1a_cdf}, \cite{Zg1a_d0}.

CDF reported the preliminary results based on a partial data set of the
1993 -- 1995 Tevatron collider run ($\sim 67 {\rm pb}^{-1}$).
The candidate events were required to have two electrons, one with
$E_T > 20$ GeV in $|\eta| < 1.1$ and another with $E_T > 20$ GeV
if $|\eta| < 1.1$,  $E_T > 15$ GeV if $1.1 < |\eta| < 2.4$ or
$E_T > 10$ GeV if $|\eta| > 2.4$, or two muons with $p_T > 20 ~{\rm GeV/c}$,
one in $|\eta| < 0.6$ and another in $|\eta| < 1.2$, and
a photon with $E_T > 7$ GeV.
The separation in $\eta - \phi$ space between a photon and a lepton
(${\cal R}_{\ell\gamma}$) had to be greater than 0.7.
This requirement suppressed the contribution of the radiative $Z$ decay
process, as in the $W\gamma$ analysis.
A total of 18 $e\nu\gamma$ and 13 $\mu\nu\gamma$ candidate events were observed.

D{\O} completed the analysis of the data set of the 1993 -- 1995 run,
and the results from the two Tevatron runs were combined.
The total data sample
corresponds to an integrated luminosity of $103~{\rm pb}^{-1}$.
For the electron channel, the candidate events were required to have two
electrons with
$E_{T} > 20 ~{\rm GeV}$ in the fiducial region of $|\eta| < 1.1$ or
$1.5 < |\eta| < 2.5$.
For the muon channel, the candidate events were required to have two muons, one
with $p_T > 15 ~{\rm GeV/c}$ in the fiducial region of $|\eta| < 1.0$ and
another with $p_T > 8(10)$ GeV/c in the fiducial region of $|\eta| < 1.1(2.4)$
for the 1992 -- 1993(1993 -- 1995) data samples.
The muon $\eta$ coverage for the 1993 -- 1995 data was extended by a track
finding method using the calorimeter hits.
The requirement for the photon was common to both the channels.
The candidate events were required to have a photon with 
$E_{T} > 10 ~{\rm GeV}$ in the fiducial region
of $|\eta| < 1.1$ or $1.5 < |\eta| < 2.5$.
The same cut as CDF on the separation between the lepton and the photon was
applied.
The above selection criteria yielded 18 $Z(ee)\gamma$ and 17 $Z(\mu\mu)\gamma$
candidates.

The backgrounds were estimated from Monte Carlo simulation and data.
The estimated total backgrounds are listed in Tables~\ref{table:Zg_cdf} and
\ref{table:Zg_d0}.
The detection efficiency was estimated
as a function of anomalous couplings
using the Monte Carlo program of Baur and Berger\cite{Zcoupling} and
a fast detector simulation program.
Form factors with a scale $\Lambda=0.5 ~{\rm TeV}$ were used in the
Monte Carlo event generation.

\begin{table}[htb]
\begin{center}
\begin{tabular}{ccc}
&
\multicolumn{2}{c}{CDF ($\int {\cal L}dt \sim 67~{\rm pb}^{-1}$)}\\ \hline
& $ee\gamma$& $\mu\mu\gamma$\\ \hline
${\rm N}_{\rm data}$& 18& 13\\
${\rm N}_{\rm BG}$& $0.9 \pm 0.3$& $0.5 \pm 0.1$\\
${\rm N}_{\rm Signal}$& $17.1 \pm 5.7$& $12.5 \pm 3.6$\\
\end{tabular}
\caption{Summary of CDF $Z\gamma\rightarrow ee\gamma, \mu\mu\gamma$ analysis}
\label{table:Zg_cdf}
\end{center}
\end{table}

\begin{table}[htb]
\begin{center}
\begin{tabular}{ccccc}
&
\multicolumn{2}{c}{D{\O} 1a ($\int {\cal L}dt = 14~{\rm pb}^{-1}$)}&
\multicolumn{2}{c}{D{\O} 1b ($\int {\cal L}dt = 89~{\rm pb}^{-1}$)}
 \\ \hline
&
$ee\gamma$& $\mu\mu\gamma$&
$ee\gamma$& $\mu\mu\gamma$\\ \hline
${\rm N}_{\rm data}$& 4& 2& 14& 15\\
${\rm N}_{\rm BG}$& $0.43 \pm 0.06$&
$0.05 \pm 0.01$& $1.8 \pm 0.6$& $3.6 \pm 0.8$\\
${\rm N}_{\rm Signal}$&
$3.6_{-1.9}^{+3.1}$& $1.9_{-1.3}^{+2.6}$& $12.1 \pm 1.2$& $17.3 \pm 2.0$\\
\end{tabular}
\caption{Summary of D{\O} $Z\gamma\rightarrow ee\gamma, \mu\mu\gamma$ analysis}
\label{table:Zg_d0}
\end{center}
\end{table}

To set limits on the anomalous couplings, the observed $E_T$
spectrum of the photon was fitted using a maximum likelihood method.
The preliminary $95\%$ CL limits on the anomalous couplings are listed in
Table~\ref{table:Zg_limits}.

\begin{table}[htb]
\begin{center}
\begin{tabular}{ccc}
& $h_{40}^Z (h_{20}^Z) = 0$& $h_{30}^Z (h_{10}^Z) = 0$\\ \hline
D{\O} 1a& $-1.9 < h_{30}^Z (h_{10}^Z) < 1.8$&
$-0.5 < h_{40}^Z (h_{20}^Z) < 0.5$\\
D{\O} 1b preliminary& $-1.3 < h_{30}^Z (h_{10}^Z) < 1.3$&
$-0.26 < h_{40}^Z (h_{20}^Z) < 0.26$\\
CDF preliminary& $-1.6 < h_{30}^Z (h_{10}^Z) < 1.6$&
$-0.4 < h_{40}^Z (h_{20}^Z) < 0.4$\\ 
\end{tabular}
\caption{$95~\%$ CL limits on $ZZ\gamma$ and $Z\gamma\gamma$ couplings}
\label{table:Zg_limits}
\end{center}
\end{table}

\subsection*{$Z\gamma\rightarrow\nu\nu\gamma$}

D{\O} completed an analysis of $Z\gamma\rightarrow\nu\nu\gamma$ process
using the 1992 -- 1993 data sample, taking advantage of its hermetic
calorimeters\cite{Znng1a_d0}.
This process has a significantly higher branching ratio than the charged lepton
decay modes of $Z$ boson and no contributions from the radiative process of
the final state leptons.
The major sources of background are $W\rightarrow e\nu$ decay with the electron
misidentified as a photon and the bremsstrahlung photon from the cosmic or
Tevatron beam halo muons.
The candidate events were required to have a photon with $E_T > 40$ GeV
in the fiducial region of $|\eta| < 1.1$ or $1.5 < |\eta| < 2.5$.
This high $E_T$ cut eliminated most of the $W\rightarrow e\nu$ background.
The $Z\rightarrow\nu\nu$ decay was identified by $\MET > 40$ GeV.
The signal to background ratio for this process is 1 to 3 with the above cuts.
D{\O} observed four candidate events.
The numbers of candidate events, background estimates and the SM prediction
are listed in Table~\ref{table:Znng}. 

\begin{table}[htb]
\begin{center}
\begin{tabular}{cc}
\multicolumn{2}{c}{D{\O} 1a ($\int{\cal L}dt = 13.5~{\rm pb}^{-1}$)}\\ \hline
${\rm N_{candidate}}$& 4\\ \hline
${\rm Muon ~background}$& $1.8 \pm 0.6$\\
$W\rightarrow e\nu ~{\rm background}$& $4.0 \pm 0.8$\\
$jj + j\gamma ~{\rm background}$& $< 0.6$\\ \hline
${\rm Total ~background}$& $5.8 \pm 1.0$\\ \hline
${\rm N_{SM}}$& $1.8 \pm 0.2$\\
\end{tabular}
\caption{Summary of D{\O} $Z\gamma\rightarrow\nu\nu\gamma$ analysis}
\label{table:Znng}
\end{center}
\end{table}

To set limits on the anomalous couplings, the observed $E_T$
spectrum of the photon was fitted using a maximum likelihood method.
The $95\%$ CL limits on the anomalous couplings are listed in
Table~\ref{table:Znng_limits} for $\Lambda = 0.5$ TeV and $\Lambda = 0.75$ TeV.
The limit contours are shown in Fig.~\ref{fig:znng_contours}.
These are the tightest limits to date among the $Z\gamma$ analyses
and the limits on $h_{40}^V$ are
better than those expected from LEP II experiments.

\begin{table}[htb]
\begin{center}
\begin{tabular}{ccccc}
& $h_{40}^Z = 0$& $h_{30}^Z = 0$& $h_{40}^{\gamma} = 0$&
 $h_{30}^{\gamma} = 0$\\ \hline
&
\multicolumn{4}{c}{$\Lambda = 0.5$ TeV}\\ \hline
$\nu\nu$& $|h_{30}^Z| < 0.87$& $|h_{40}^Z| < 0.21$&
 $|h_{30}^{\gamma}| < 0.90$& $|h_{40}^{\gamma}| < 0.22$\\
1a $(ee,\mu\mu,\nu\nu)$& $|h_{30}^Z| < 0.78$& $|h_{40}^Z| < 0.19$&
 $|h_{30}^{\gamma}| < 0.90$& $|h_{40}^{\gamma}| < 0.22$\\ \hline
&
\multicolumn{4}{c}{$\Lambda = 0.75$ TeV}\\ \hline
$\nu\nu$& $|h_{30}^Z| < 0.49$& $|h_{40}^Z| < 0.07$&
 $|h_{30}^{\gamma}| < 0.50$& $|h_{40}^{\gamma}| < 0.07$\\
1a $(ee,\mu\mu,\nu\nu)$& $|h_{30}^Z| < 0.44$&
$|h_{40}^Z| < 0.06$&
$|h_{30}^{\gamma}| < 0.45$&
$|h_{40}^{\gamma}| < 0.06$\\
\end{tabular}
\caption{D{\O} $95~\%$ CL limits on $ZZ\gamma$ and $Z\gamma\gamma$ couplings}
\label{table:Znng_limits}
\end{center}
\end{table}

\begin{figure}[htb]
 \centerline{\includegraphics[width=9cm]{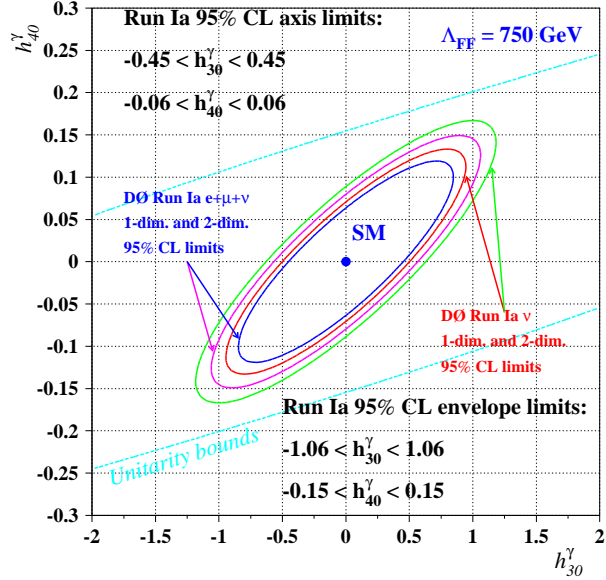}}
 \caption{$95~\%$ CL limits on the anomalous $Z\gamma\gamma$ couplings from
 $Z\gamma\rightarrow\nu\nu\gamma$ analysis for $\Lambda = 0.75$ TeV.}
 \label{fig:znng_contours}
\end{figure}

\section*{Summary}

The CDF and D{\O} experiments at Fermilab set limits on anomalous trilinear
gauge boson couplings using four diboson final states,
$W\gamma\rightarrow e\nu\gamma, \mu\nu\gamma$,
$WW\rightarrow$ dileptons, $WW/WZ\rightarrow \ell\nu jj, \ell\ell jj$
and $Z\gamma\rightarrow ee\gamma, \mu\mu\gamma, \nu\nu\gamma$.
The tightest limits on the anomalous $WW\gamma$ couplings with no assumptions on
the $WWZ$ couplings were obtained from the D{\O} $W\gamma$ analysis:
\begin{eqnarray*}
-0.98 < \Delta\kappa_{\gamma} < -0.94 ~(\lambda_{\gamma} = 0)~;~
-0.31 < \lambda_{\gamma} < 0.29 ~(\Delta\kappa_{\gamma} = 0).
\end{eqnarray*}
The tightest limits on the anomalous $WW\gamma$ and $WWZ$ couplings, with the assumption
that the two sets of couplings are equal, were obtained from a combined fit
to $W\gamma$, $WW\rightarrow$ dileptons and $WW/WZ\rightarrow e\nu jj$ data
samples by D{\O}:
\begin{eqnarray*}
-0.33 < \Delta\kappa < -0.45 ~(\lambda = 0)~;~
-0.20 < \lambda < 0.20 ~(\Delta\kappa = 0).
\end{eqnarray*}
CDF measured the production cross section of $W$ boson pair using dilepton
decay modes:
\begin{eqnarray*}
\sigma({\bar p}p\rightarrow W^+W^-) &=& 10.2_{-5.1}^{+6.3}~({\rm stat}) \pm
1.6~({\rm syst})~{\rm pb}.
\end{eqnarray*}
The tightest limits on the anomalous $ZZ\gamma$ and $Z\gamma\gamma$ couplings
were obtained from a $Z\gamma\rightarrow\nu\nu\gamma$ analysis by D{\O} using
the 1992 -- 1993 Tevatron collider run data:
\begin{eqnarray*}
|h_{30}^Z| < 0.44 ~(|h_{40}^Z| = 0)&;&~|h_{40}^Z| < 0.06 ~(|h_{30}^Z| = 0)\\
|h_{30}^{\gamma}| < 0.45 ~(|h_{40}^{\gamma}| = 0)&;&
~|h_{40}^{\gamma}| < 0.06 ~(|h_{30}^{\gamma}| = 0).
\end{eqnarray*}
The CDF and D{\O} experiments plan to combine the limits on the anomalous trilinear
gauge boson couplings
and improve the limits on the couplings.


\end{document}